\documentclass[10pt, conference]{IEEEtran} 
\usepackage{caption}

\usepackage[top=0.65in, bottom=1.02in, left=0.64in, right=0.64in]{geometry}
\IEEEoverridecommandlockouts

\usepackage{subcaption}
\usepackage{amsmath,mathtools}
\usepackage{amsfonts}
\usepackage{amssymb}
\usepackage{amsthm}
\usepackage{algorithm}
\usepackage{algpseudocode}
\usepackage{booktabs} 
\usepackage{fancyhdr}
\usepackage{xcolor}
\usepackage{multicol}
\usepackage{graphicx}
\usepackage{caption}
\usepackage{subcaption}  
\usepackage{soul}
\usepackage{pifont}
\usepackage[breaklinks=true]{hyperref}
\usepackage{boxedminipage}
\usepackage{tikz}
\usepackage{multirow}
\usepackage{array}
\usepackage{enumitem,kantlipsum}
\usepackage{comment}
\usepackage{tikz}
\usetikzlibrary{patterns}
\usepackage{tikzscale}
\newtheorem{theorem}{Motivation}

\usepackage{url}

\begin{document}

\title{\LARGE{ACCORD: \underline{A}pplication \underline{C}ontext-aware \underline{C}ross-layer \underline{O}ptimization and \underline{R}esource \underline{D}esign for 5G/NextG Machine-centric Applications}}
\author{\IEEEauthorblockN{Azuka Chiejina\IEEEauthorrefmark{1}, Subhramoy Mohanti\IEEEauthorrefmark{2}, and Vijay K. Shah\IEEEauthorrefmark{1}}
\IEEEauthorblockA{\IEEEauthorrefmark{1}NC State University, USA, \IEEEauthorrefmark{2}InterDigital Communications, Inc., USA}
\IEEEauthorblockA{Email: ajchieji@ncsu.edu, subhramoy.mohanti@interdigital.com, vijay.shah@ncsu.edu}}
\maketitle
\begin{abstract}
Recent advancements in artificial intelligence (AI) and edge computing have accelerated the development of machine-centric applications (MCAs), such as smart surveillance systems. In these applications, video cameras and sensors offload inference tasks like license plate recognition and vehicle tracking to remote servers due to local computing and energy constraints. However, legacy network solutions, designed primarily for human-centric applications, struggle to reliably support these MCAs, which demand heterogeneous and fluctuating quality of service (QoS) (due to diverse application inference tasks), further challenged by dynamic wireless network conditions and limited spectrum resources. To tackle these challenges, we propose an Application Context-aware Cross-layer Optimization and Resource Design (ACCORD) framework. This innovative framework anticipates the evolving demands of MCAs in real time, quickly adapting to provide customized QoS and optimal performance, even for the most dynamic and unpredictable MCAs. This also leads to improved network resource management and spectrum utilization. ACCORD operates as a closed feedback-loop system between the application client and network and consists of two key components: (1) Building Application Context: It focuses on understanding the specific context of MCA requirements. Contextual factors include device capabilities, user behavior (e.g., mobility speed), and network channel conditions. (2) Cross-layer Network Parameter Configuration: Utilizing a deep reinforcement learning (DRL) approach, this component leverages the contextual information to optimize network configuration parameters across various layers, including physical (PHY), medium access control (MAC), and radio link control (RLC), as well as the application layer, to meet the desired QoS requirement in real-time. Extensive evaluation with the 3GPP-compliant MATLAB 5G toolbox demonstrates the practicality and effectiveness of our proposed ACCORD framework.
\end{abstract}
\section{Introduction}
The continuous evolution of cellular networks has enabled unprecedented connectivity, characterized by faster speeds and lower latency. While initially designed for human-centric applications like video streaming, the horizon of network usage is rapidly expanding. Recent advancements in artificial intelligence (AI) have fueled the growth of machine-centric applications (MCAs), such as object detection and tracking, where machines rely on inferences derived from sensor data (e.g., video) to perform machine tasks like surveillance and navigation. These MCAs are transforming emerging use-cases such as connected vehicles\,\cite{V2X-DoT}, smart surveillance\,\cite{akyildiz2022wireless}, and automated factories\,\cite{bai2020industry}. Existing research has focused on enhancing the communication architecture of MCAs, enabling resource-constrained end-devices to offload their sensor data to powerful remote servers for complex inference processing and generating feedback for executing the machine task. This strategy, known as {\em remote inferencing}\,\cite{matsubara2022split}, effectively overcomes the limitations of individual end-devices, and has led to the rise of machine-type communications (MTC). However, the unique Quality of Service (QoS) requirements of MTC present significant challenges for 5G networks.\newline
\begin{figure}[t!]
\centering
\includegraphics[trim=20 15 20 12,clip,width=\linewidth]{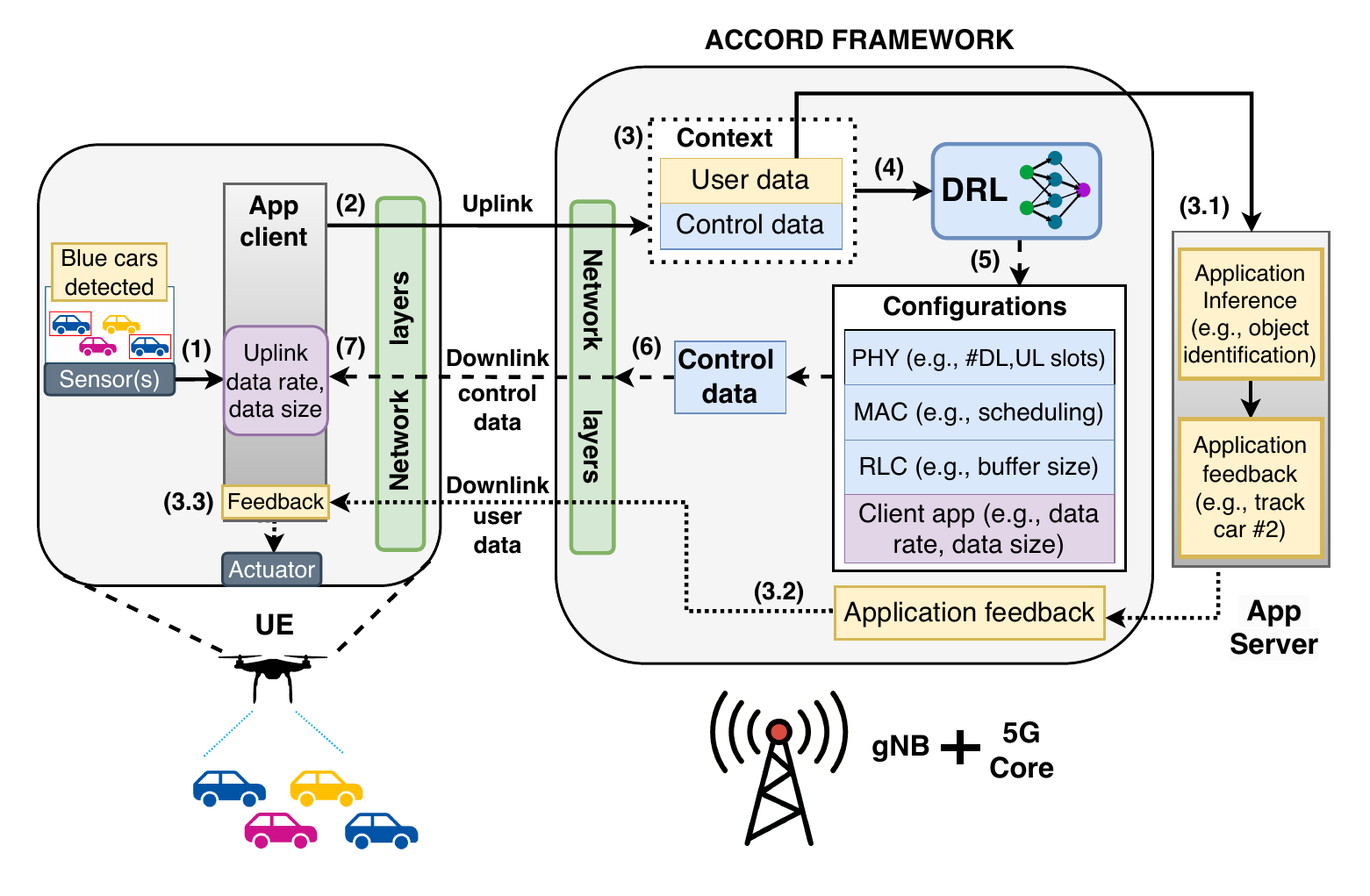}
\caption{ACCORD framework in 5G. The application data (yellow) and network control data (blue) provide the context that is used as input for the deep reinforcement learning (DRL) to generate the configurations for the network (blue) and application (magenta) for machine-type communication.} 
\label{fig:arch}
\end{figure}
\begin{figure*}[t!]
\centering
\includegraphics[trim=0 0 0 0,clip,width=\linewidth]{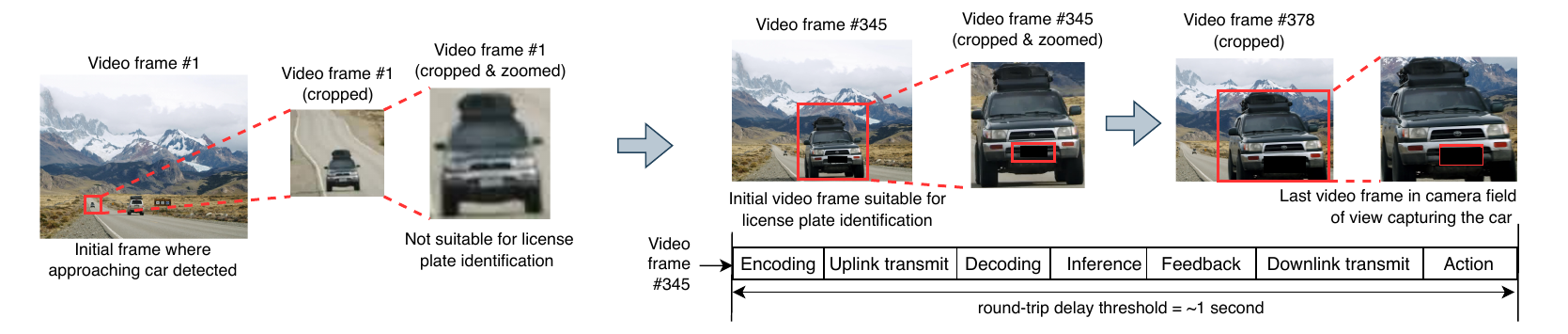}
\caption{Example scenario for estimating the threshold of round-trip time of video frame generation to inference feedback reception from remote server for a machine-centric task such as smart surveillance.} 
\vspace{-3mm}
\label{fig:app_charac}
\end{figure*}
\noindent $\bullet$ \textbf {Challenges for 5G in serving MCAs:} Human-centric applications like video streaming or online gaming have predictable needs for throughput and latency. They are mostly downlink-heavy\,\cite{ghoshal2022depth}\,\cite{fezeu2023depth}, and have predictable traffic patterns that allow statistical modeling\,\cite{rao2011network}. The QoS in 5G is managed by grouping applications into {\em classes}\,\cite{5QI}, with fixed upper and lower bounds of the Key Performance Indicators (KPI) like latency, based on the similarity of the corresponding statistical traffic models. 
In contrast, MCAs are mostly uplink-heavy and are often triggered by {\em events} in the real world, leading to sudden {\em bursts} of {\em time-sensitive} data that leads to {\em unpredictable} QoS demands\,\cite{liu2019edge}\,\cite{feng2021ultra}\,\cite{kong2017object}.
An MCA using video for smart surveillance through remote inferencing in a sparse environment (e.g., rural highway) might have very different QoS needs than the same MCA in a dense environment with fast-moving objects (e.g., crowded urban intersection). For 5G to provide {\em guaranteed} service to MCAs with dynamic upper and lower bounds in the KPIs, the network has to over-allocate resources for each MCA to meet the unpredictable demands. This can overwhelm the limited spectrum and is not scalable.
Manually configuring network parameters for each MCA scenario is also impractical due to the sheer number of parameters and the complexity of real-world situations. \newline
\noindent \textbf {Proposed Work:} 
To overcome the limitations of legacy 5G networks in handling the unique demands of MCAs, we propose an \underline{A}pplication \underline{C}ontext-aware \underline{C}ross-layer \underline{O}ptimization and \underline{R}esource \underline{D}esign (ACCORD) framework in 5G, that is depicted in Fig.\,\ref{fig:arch}. ACCORD understands the dynamic demands of MCAs by {\em learning} the application {\em context}.
This context information includes the application latency requirement, observed latency, device configuration, user mobility, environmental dynamics, and observed network conditions. This information is used as input state representation to a Deep Reinforcement Learning (DRL) model that learns how to meet the application latency requirement in a spectrum-efficient manner by observing the change in the context after applying different configurations of the network layers (PHY, MAC, RLC) and the application layer. 
For practical implementation of this framework, the RAN must be capable of exposing real-time data, providing analytics, and supporting closed-loop control. While traditional 5G networks lack these functionalities, the advent of Open RAN paradigm has enabled the design of NextG networks in a modular and disaggregated fashion\,\cite{polese2023understanding}, with open and standardized interfaces that enable access to the necessary data and analytics for implementing frameworks like ACCORD. \newline
The main contributions of this paper are as follows:
\begin{itemize}
    \item We perform experimental evaluations of an example generalizable MCA and characterize its requirements in a real-world scenario. We showcase how MCA requirements are driven by the triggering event characteristics, the environment, the behavior of the user (e.g mobility) and the objects in the environment, thereby making these requirements unpredictable.
    \item We leverage 3GPP-compliant Matlab 5G toolbox\,\cite{matlab}, to run data communication experiments with an example generalizable MCA. We characterize the performance limitations of legacy 5G network optimizations in the PHY, MAC, RLC, and application (APP) layers when serving MCAs with dynamic requirements.  
    \item We explain the building blocks of ACCORD, and through extensive experiments we showcase its performance in a 5G environment, serving MCAs in different scenarios. We show how our proposed approach ensures the MCA QoS through better resource management and spectrum utilization, compared to legacy 5G network solutions.
    \end{itemize}
\begin{figure*}[t!]
    \centering
    \begin{subfigure}[b]{0.24\textwidth} 
        \centering
        \includegraphics[width=\linewidth]{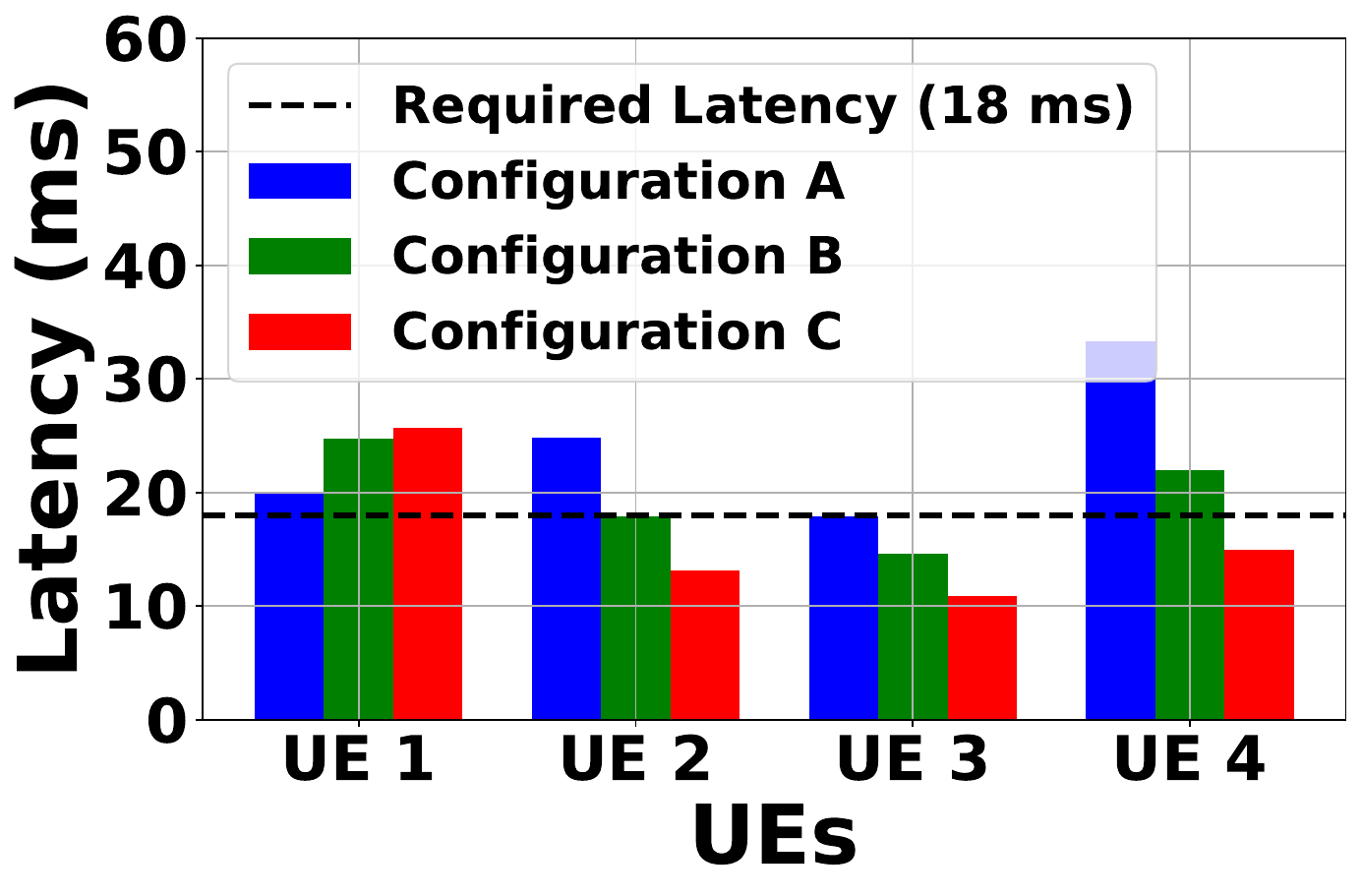}
        \caption{Profile\,1}
        \label{fig:prelim_images_a}
    \end{subfigure}
    \hfill 
    \begin{subfigure}[b]{0.24\textwidth}
        \centering
        \includegraphics[width=\linewidth]{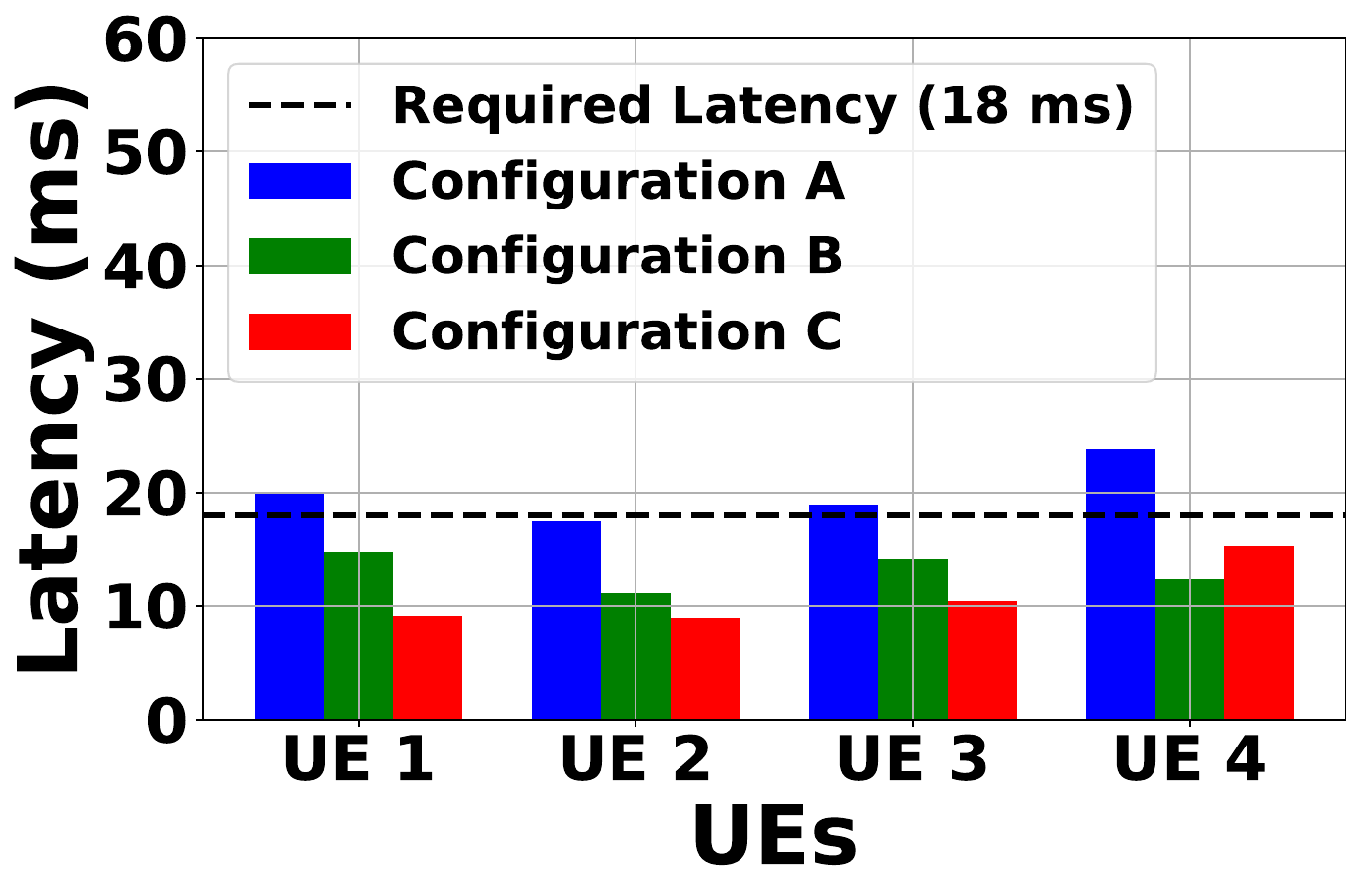}
        \caption{Profile\,2}
        \label{fig:prelim_images_b}
    \end{subfigure}
    \hfill
    \begin{subfigure}[b]{0.24\textwidth}
        \centering
        \includegraphics[width=\linewidth]{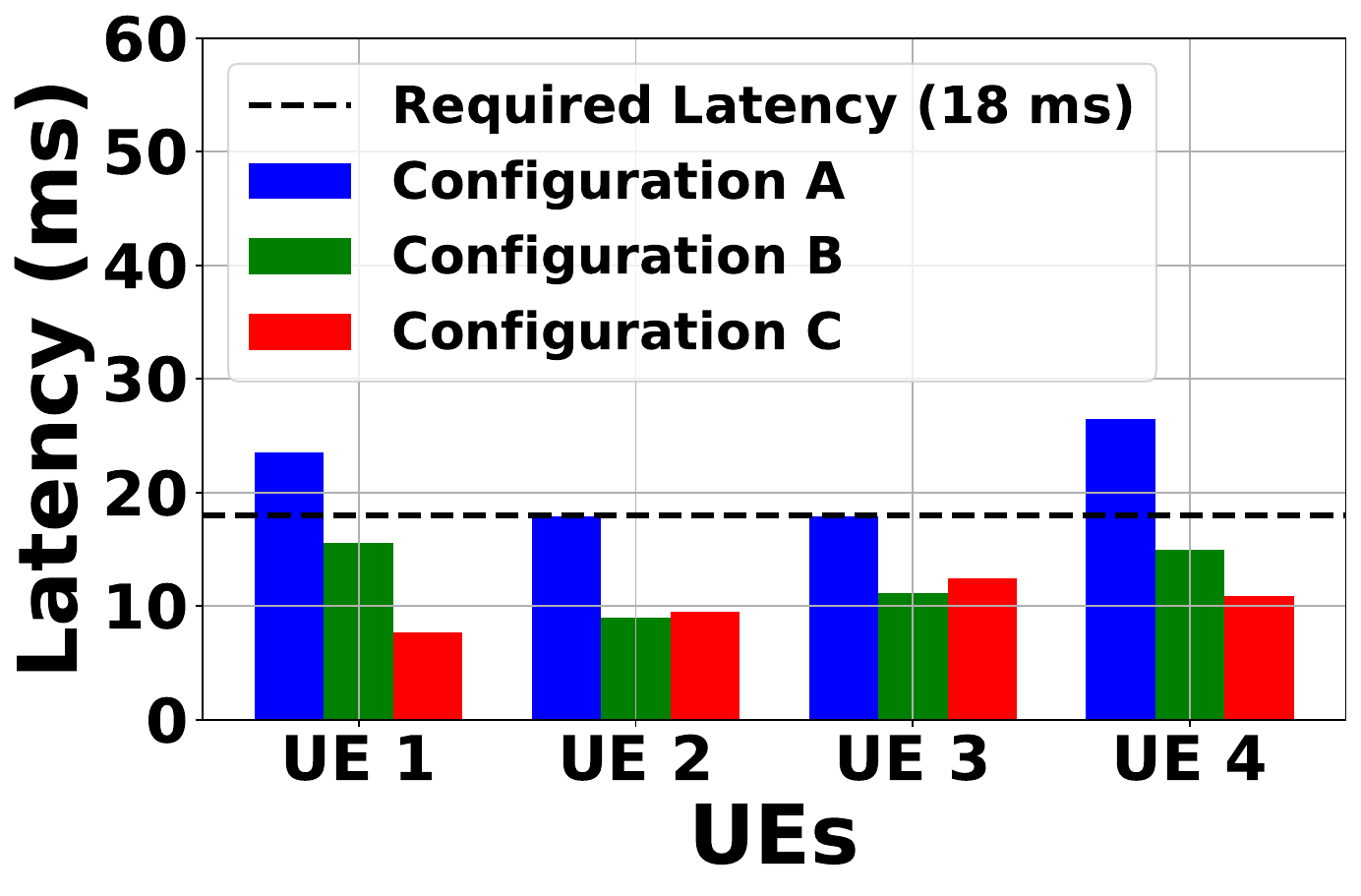}
        \caption{Profile\,3}
        \label{fig:prelim_images_c}
    \end{subfigure}
    \hfill
    \begin{subfigure}[b]{0.24\textwidth}
        \centering
        \includegraphics[width=\linewidth]{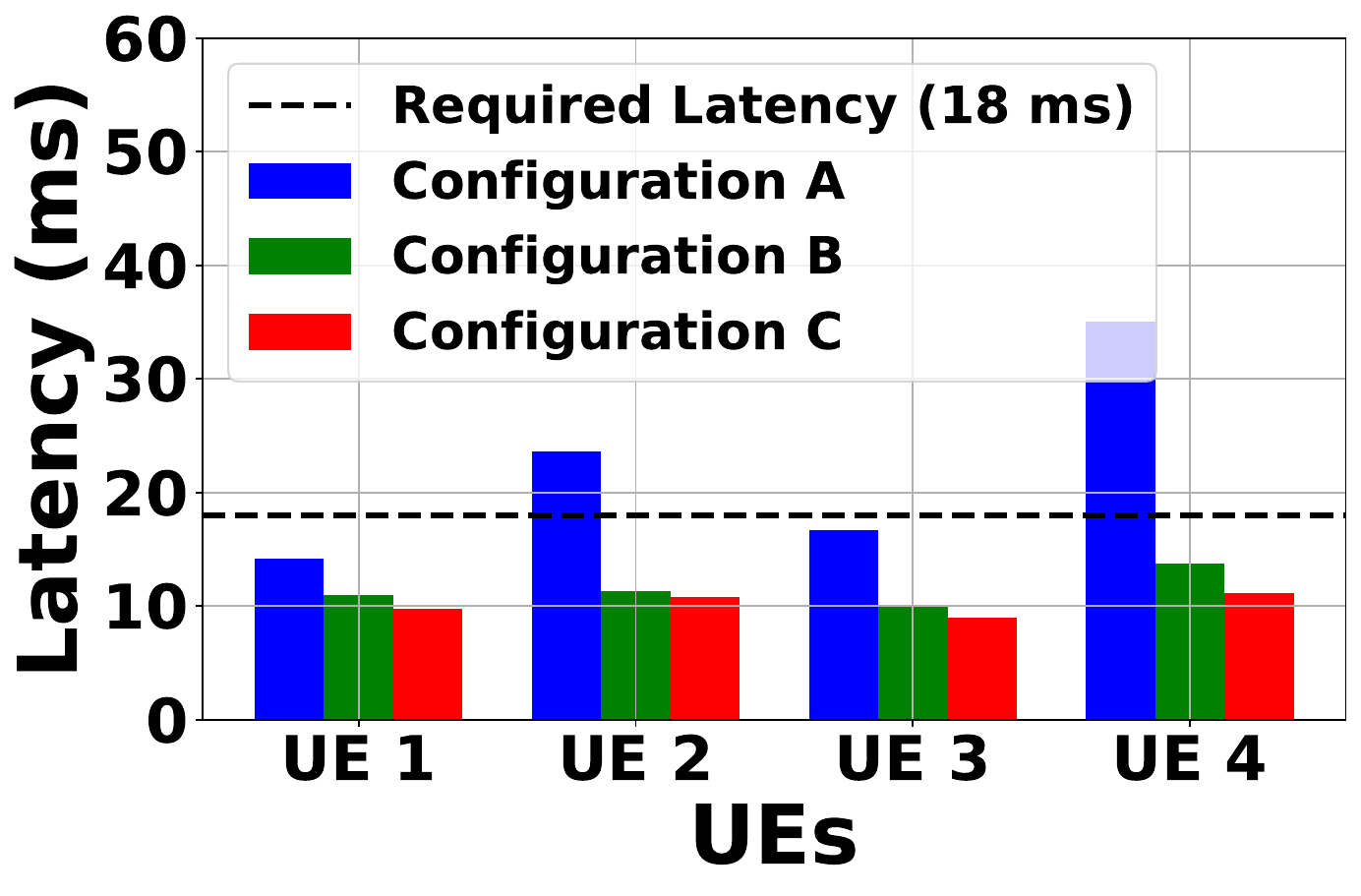}
        \caption{Profile\,4}
        \label{fig:prelim_images_d}
    \end{subfigure}
    \caption{Impact of legacy 5G configurations on achieving a target latency of 18\,ms for 4 UEs at varying distances from the gNB.}
    \label{fig:prelim_images}
    \vspace{-2mm} 
\end{figure*}

\section{Related Work}
Optimizing Radio Access Network (RAN) performance through network-layer enhancements, particularly in Time Division Duplex (TDD) slot configurations, has been extensively studied. Static TDD configurations are inherently inefficient as they cannot adapt to real-time network traffic variations. To address this, various dynamic TDD techniques leveraging deep reinforcement learning (DRL) have been proposed to optimize throughput and latency. However, these approaches face challenges related to scalability and the need for real-time decision-making.

Several works\,\cite{bagaa2021using, boutiba2023enabling, hassan2024wixor, boutiba2023multi, park2024deep, sun2015d2d, dao2021deep, ghoshal2024enabling, tang2020deep} explore DRL-driven dynamic TDD solutions across different scenarios, including high-density environments, device-to-device (D2D) communications, massive IoT, and public/private 5G networks. In particular,\,\cite{bagaa2021using, boutiba2023enabling} introduce a TDD xApp that leverages buffer status reports (BSR) and downlink buffer size to prevent buffer overflow. Similarly,\,\cite{hassan2024wixor} enhances user experience (QoE) by dynamically configuring TDD slots and symbols, utilizing both BSR and channel quality indicators (CQI) for adaptive modulation and coding schemes (MCS). This method operates at the radio protocol layer, making it application-transparent. Meanwhile,\,\cite{ghoshal2024enabling} employs UE-side wireless channel metrics for proactive PHY frame configuration.

While most research primarily focuses on PHY-layer optimization, some studies explore higher-layer enhancements. MAC scheduling optimization has been examined in\,\cite{yan2019intelligent, chen2024elase}, while \cite{kumar2018dynamic} introduces a dynamic RLC buffer sizing algorithm that minimizes latency by adjusting buffer size based on real-time queuing delay measurements. Additionally,\,\cite{fezeu2024roaming} provides a comprehensive cross-layer measurement study, linking PHY, MAC, and RLC configurations to their impact on overall network performance.

A similar adaptive approach has long been applied in internet and video streaming domains, where adaptive bitrate streaming techniques assess network conditions and adjust video quality accordingly. These techniques involve selecting the appropriate resolution, modifying frame rates, and optimizing encoding strategies\,\cite{nam2014youslow, zuo2022adaptive, tran2020bitrate}. Such methodologies have been widely adopted in modern streaming platforms like Netflix and YouTube.

Building on these foundations, our work characterizes MCA requirements in real-world scenarios and develops a multi-layer optimization framework spanning PHY, MAC, and RLC layers. Our approach ensures adaptive resource allocation under diverse mobility and channel conditions, optimizing performance with minimal resource overhead.

\section{Preliminary Experiments}
\label{sec:preliminary_experiment}
\subsection{Characterizing MCA requirement:}
MCAs often exhibit dynamic and unpredictable behavior, influenced by external factors such as user actions and environmental conditions.
Consider the example shown in Fig.\,\ref{fig:app_charac}, where a vehicle is approaching a mobile monitoring device equipped with a camera. This device captures video frames and transmits them to a remote server for license plate identification. The time required for the vehicle to be detected and identified, and for the feedback to be relayed back to the device (e.g., ``track vehicle") determines the necessary round-trip latency of the application. This latency is affected by various factors, including the vehicle speed, the camera field of view (FoV), and the processing time required for identification. In this particular scenario, the round-trip latency is $\sim$1\,second, (including uplink latency), as the vehicle remains within the camera field of view for 33 frames at a rate of 30 frames per second (FPS). If the vehicle's speed were to change or the camera field of view were different, the required latency would also change. 
This example highlights the dynamic nature of MCAs and the need for network optimization strategies that can adapt to their unique and changing requirements.
\vspace{-0.05in}
\subsection{Network Characterization:}
We conduct simulation experiments with an MCA (similar to Sec.\,\ref{sec:preliminary_experiment}.A) over a 5G network to understand the uplink performance as the MCA transmit video to a 5G network-connected server.
We track the application data flow through the network stack and quantify the limitations of using standard-compliant fixed configurations across the PHY, MAC, RLC, and APP layers\,\cite{fezeu:techreport2023}\,\cite{3gppmac}\,\cite{3gpprlc}. 
\begin{table}[t!]
\vspace{-1mm}
 \caption{Configuration Profiles}
 \vspace{-1mm}
 \begin{center}
 \resizebox{0.482\textwidth}{!}{
  \begin{tabular}{|c|c|c|c|}
 \hline
\textbf{Profile}  & \textbf{Frame Rate (FPS)} & \textbf{MAC scheduling}  & \textbf{RLC buffer (KB)}\\
\hline
 1 & 30 & Round Robin & 6 \\
 \hline
 2 & 60 & Round Robin & 6 \\
 \hline
 3 & 60 & Round Robin & 10 \\
 \hline
 4 & 60 & Proportional Fair & 6 \\
 \hline
 \end{tabular}}
 \end{center}
 \label{tab:notations}
 \vspace{-6mm}
\end{table}

\noindent $\bullet$ \textbf {Experiment Setup:} We use a 3GPP-compliant MATLAB 5G toolbox featuring a base station (gNB) having an 8x8 MIMO configuration with a bandwidth (BW) of 10\,MHz with 30\,KHz subcarrier spacing (SCS), and 4 UEs each having 2x2 MIMO configuration at different locations from the gNB (300\,m, 1200\,m, 1500\,m, 3000\,m) and experiencing channel conditions based on the 3GPP standard clustered delay line (CDL) channel model. 
The MCA on each UE generated uplink video frames of size 7.5\,KB, at rates varying between 30\,FPS and 60\,FPS. The MCA requirement was an uplink latency of 18\,ms based on the vehicle speed and camera FoV, calculated based on a scenario similar to the one described in Sec.\,\ref{sec:preliminary_experiment}.A. 
We consider three different PHY slot configurations that are downlink-heavy with 7\,downlink (DL)-3\,uplink (UL) slots \texttt{(Configuration A)}, equalized with 5\,DL-5\,UL slots \texttt{(Configuration B)}, and an uplink-heavy configuration with 3\,DL-7\,UL slots \texttt{(Configuration C)}, each with a periodicity of 5\,ms. Table\,\ref{tab:notations} highlights the considered baseline 3GPP parameters across the other layers. 
\noindent $\bullet$ \textbf {Observation:} The experimental results presented in Fig.\,\ref{fig:prelim_images} illustrate the impact of various 5G layer configurations on achieving the \textcolor{black}{target latency of 18\,ms} for 4\,UEs situated at different distances from the gNB. In Profile\,1, (Fig.\,\ref{fig:prelim_images_a}), modifying only the PHY parameters, specifically the DL and UL slot allocations, resulted in latency improvements for UEs\,2,\,3,\,and\,4, while UE\,1 experienced no significant change. This suggests that PHY layer adjustments alone may not guarantee latency requirements for UEs experiencing different channel conditions.
Profiles\,2,\,3,\,and\,4 (Fig.\,\ref{fig:prelim_images_b},\,\ref{fig:prelim_images_c},\,\,\ref{fig:prelim_images_d}) demonstrate the impact of adjusting parameters across PHY, MAC, RLC, and APP layers together in meeting the required latency under different channel conditions. These configurations yielded more consistent latency improvements across all the UEs for specific configuration profiles.

{\em Our observations highlight the limitations of relying solely on fixed configurations or isolated layer-specific optimizations in 5G to meet diverse MCA requirements, especially in varying channel conditions.
This motivates the need for exploring optimization strategies that consider the interplay between network and application layers, enabling dynamic adjustments tailored to individual MCA needs. }
\section{Accord framework description}
\label{sec:solution}
\vspace{0.05in}
To enable the cross-network and application layer optimization, we develop and implement the ACCORD framework over a 5G network. We first provide a detailed workflow of the framework, referring to Fig.\,\ref{fig:arch}, and then explain the details of the DRL solution that resides at the core of the framework.
\subsection{ACCORD workflow}
As the MCA client-server communication instantiates over 5G, ACCORD starts to optimize the network and application through the following steps:

\noindent $\bullet$ \textbf{Step 1-2: } An MCA client, running on the UE, generates video frames (user data) by observing events in the environment (e.g., vehicle detection). After characterizing the requirement (explained in Sec.\,\ref{sec:preliminary_experiment}.A), the MCA client transmits the user data to the server using legacy network configuration parameters. The MCA client also transmits the calculated requirement along with information about UE buffer space and channel quality indicator (CQI) through uplink control information (UCI) messages to the gNB.

\noindent $\bullet$ \textbf{Step 3 -  Building MCA context: } After receiving the user and control data from the UE, ACCORD combines this information with network control data (UE mobility/position and MCS), to build the context for MCA requirement.

\noindent $\bullet$ \textbf{Step 4-5:} The context information is used as input to the DRL agent to perform cross-layer optimization by selecting the optimal set of configurations across the network and application layers, that meet the MCA requirement.

\noindent $\bullet$ \textbf{Step 6-7:} The lower network layer configurations (PHY, MAC) from ACCORD are implemented at the network side (e.g., gNB). The RLC and APP configurations are transmitted to the UE through the downlink control information (DCI).

In \texttt{Step\,3.1}, the MCA server uses the uplink data from the client (that was transmitted in Step 2) to make inferences and generates that feedback to the UE through the network as seen in \texttt{Steps\,3.2} and \texttt{3.3}.\newline
\vspace{-4mm}
\subsection{Proposed DRL approach for context awareness}
To develop the context-aware cross-network and application layer optimization in ACCORD, we train a DRL agent by making it interact with the 5G environment simulated in the 3GPP-compliant MATLAB 5G toolbox. The DRL objective is to learn the optimal configuration parameters across the PHY, MAC, RLC, and APP layers based on the MCA context. 
Through this interaction, the agent obtains a reward related to the objective function of meeting the latency requirements of the UE while using minimal network resources to save spectrum. This approach was used as opposed to the supervised learning procedure which necessitates a comprehensive dataset capturing all possible network conditions, UE behaviors, and requirements, which is impractical considering the unpredictable MCA requirement and complexity of real-world situations. The optimization problem can be expressed as a Markov decision process (MDP) of the DRL, which we describe as follows:

\noindent $\bullet$ \textbf{State: } The state space $s_{t}$ can be modeled as the combination of the application required latency ($\mathcal{L}_t$), transmitted bytes ($c_t$)), UE location ($g_t$), MAC buffer status report (BSR) ($k_t$), RLC buffer size ($p_t$), CQI ($v_t$) and MCS ($u_t$) at current time window $t$ with a window size of $w$ for all the UEs in set $U$. Overall, $s_{t}$ is formulated as $s_{t} = \{ \mathcal{L}_t, c_t, g_t, k_t, p_t, v_t, u_t\}$. The $g_t$ is a tuple of dimension 2 corresponding to the 2D coordinates of the target location.

\noindent $\bullet$ \textbf{Action: } The action space is designed to allow the DQN agent to generate the PHY, MAC, RLC and APP layer configuration parameters at time $t$ for time window $t+w$ (the next time window). We define the action space $\mathcal{A}_{t}$ for all the UEs in $U$ ($|U| = \mathcal{K}$), as the combination of $C_{t}^{PHY}$ (PHY frame configuration), $C_{t}^{MAC}$ (MAC configuration), $C_{t}^{RLC}$ (RLC configuration) and $C_{t}^{APP}$ (APP configuration), at time $t$ for time window $t+w$, hence, $\mathcal{A}_{t} =\mathcal{K} \times \{C_{t}^{PHY}, C_{t}^{MAC}, C_{t}^{RLC}, C_{t}^{RLC}\}$. The action $\mathcal{A}_{t}$ is generated by the designed policy $\pi_{\theta}({s_t})$ from the state $s_{t}$. Hence, the action of time $t$ is formulated as ${A}_{t} = \pi_{\theta}(s_{t})$.

\noindent $\bullet$ \textbf {Reward: } The reward space is a function of the action $\mathcal{A}_{t}$ generated at time $t$ and transition to the next state $s_{t+w}$ at time $t+w$ for each UE as: $f^{\mathcal{R}}_{\mathcal{A}}(\mathcal{A}_{t},s_{t+w}) = r_{t}$, where $r_{t}$ is the achieved reward for a UE at time $t$. Our reward function is defined as $r_{t}=
\left( \frac{1}{1 + e^{-\beta (l_t - \mathcal{L}_t)}} \right) \times 100$ for $l_t \leq \mathcal{L}_t$ and $r_{t}= 0$ otherwise.
$\beta$ is chosen in such a way as to control the steepness of the sigmoid function utilized in the equation. Smaller values of $\beta$ ensures that the function is smoother, causing $r_t$ to change more gradually as $l_t$ moves around $\mathcal{L}_t$.\newline

{\em Our goal is to find an optimal policy that minimizes the overall network latency based on real-time MCA requirements and network conditions while conserving network resources. This is achieved by mapping the state space to an action space that maximizes the accumulated reward. We employ the Deep Q-Network (DQN) algorithm\,\cite{mnih2013playing} explained in Algorithm\,\ref{Algo:dqn} and adapt it to solve our MDP.}
\begin{algorithm}[t!] 
\caption{Context-Aware DQN in ACCORD}
\scriptsize
\begin{algorithmic}
\State \textbf{Initialization:} Experience replay memory $\mathcal{D}$ to capacity $N$, batch size, main network parameters $\theta$, target network parameters $\theta^- = \theta$, update rate of target network $\tau$, number of episodes $N_{\text{eps}}$, number of time steps in each episode $N_{\text{step}}$, learning rate and epsilon $\epsilon$ for the exploitation and exploration.
\For{\textit{episode} = 1 : $N_{\text{eps}}$}
    \State Initialize processed state from environment $\phi_t$ =  $\phi(s_t)$;
    \For{\textit{step} = 1 : $N_{\text{step}}$}
        \State Using epsilon greedy as in \cite{mnih2013playing};
        \State With probability $\epsilon$, select a random action $a_t$
        \State Otherwise, select $a_t = \max_a Q^*(\phi_t, a; \theta)$
        \State Execute action $a_t$ and observe reward $r_t$ and $s_{t+w}$
        \State Process $s_{t+w}$ and store $(\phi_t, a_t,\phi_{t+w}, r_t)$ in $\mathcal{D}$
        \State \textbf{Optimize model:}
         \If{memory has sufficient transitions}
            \State Sample a random mini-batch of transitions $(\phi_k, a_k, \phi_{k+w},r_k)$ from $\mathcal{D}$
            \State Compute $Q(\phi_k, a_k; \theta)$ for each state-action pair in the batch
             \State Compute target Q values based on if $\phi_{k+w}$ is terminal or not as in \cite{mnih2013playing}
            \State Compute loss between $Q(\phi_k, a_k; \theta)$ and the target Q values
            \State Perform gradient descent to update $\theta$;
        \EndIf
            \State Soft update target network: $\theta^{-} \gets \tau \theta + (1 - \tau)\theta^{-}$
    \EndFor
\EndFor
\end{algorithmic}
\label{Algo:dqn}
\end{algorithm}
\section{Performance Evaluation}
We evaluate the performance of ACCORD in 5G when serving the example MCA in two distinct scenarios, each representing a practical use case with unique characteristics and challenges.
\subsection{Experiments}
\begin{figure}[t!]
    \centering
\includegraphics[trim=20 10 40 20,clip,width=\linewidth]{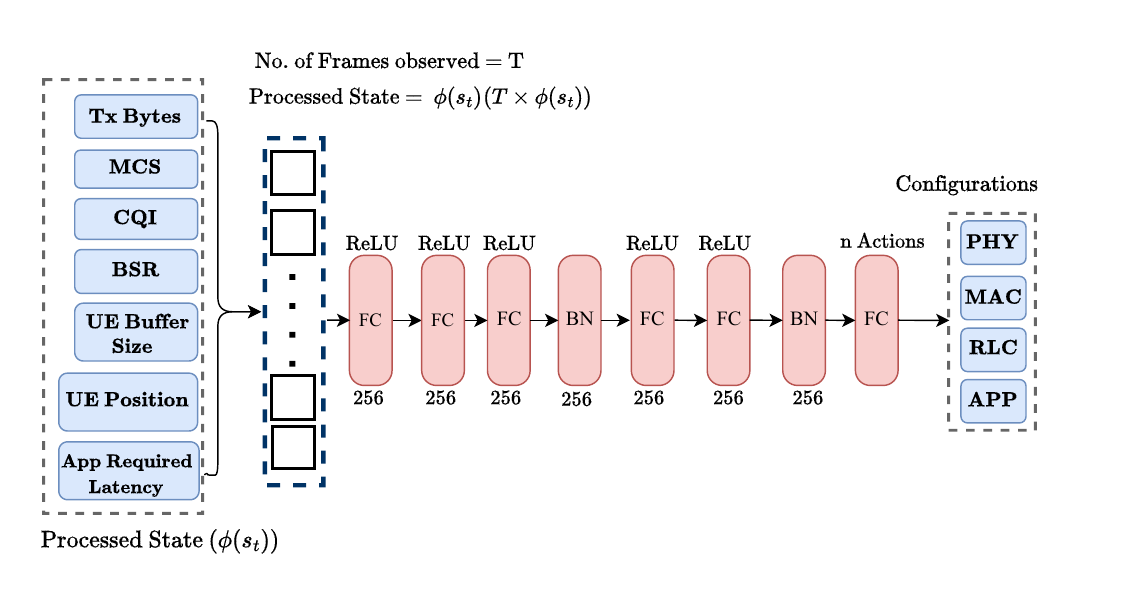}
    \vspace{-5mm}
    \caption{MLP Model used for training the DQN in ACCORD.}
    \vspace{-2mm}
\label{fig:drl_agent_architecture}
\end{figure}

The considered scenarios are simulated with gNB, UEs and MCA configurations mirroring those in the preliminary experiments (Sec.\,\ref{sec:preliminary_experiment}) but with different latency requirements in a reduced BW of 5\,MHz to show spectrum efficiency.\newline
\noindent $\bullet$ \textbf{Scenario A -  Stationary UEs: }
We consider two cases: (1) a single UE positioned 5000\,m from the gNB with a CQI of 10, and (2) two UEs at distances of 600\,m and 3600\,m, with CQIs of 15 and 8, respectively.\newline
\noindent $\bullet$ \textbf{Scenario B - Mobile UEs: }
Here also we consider two cases: (1) a single UE moving away from the gNB at 60\,mph with decreasing CQI over time, and (2) two UEs, one with degrading CQI and the other with improving CQI.\newline
In both scenarios, the gNB and UEs are initialized with random configurations across all layers.\newline
\noindent $\bullet$ \textbf{DRL Architecture \& Training: }The DRL agent in ACCORD employs a DQN algorithm described in Sec.\,\ref{sec:solution}.B and Algorithm\,\ref{Algo:dqn} with the following hyper-parameters: batch size\,128, discount factor (gamma)\,0.2, replay memory capacity\,10,000, and Adam optimizer with a learning rate\,0.0001. The neural network architecture, depicted in Fig.\,\ref{fig:drl_agent_architecture}, utilizes a multilayer perceptron with ReLU activation functions, fully connected layers (FC) with 256 neurons each, and batch normalization (BN) after specific layers. The input to the network consists of key performance metrics collected every 10\,ms that form the State space described in Sec.\,\ref{sec:solution}.B.
These metrics are aggregated over a 100\,ms window, preprocessed, and fed into the network as a vector of shape (40,1) for a single UE and (80,1) for two UEs. The network output is mapped to specific configurations across the PHY, MAC, RLC, and APP layers using a predefined codebook. Based on the selected action, the reward is generated using the reward function described in Sec.\,\ref{sec:solution}.B.

\begin{figure}[h!] 
    \centering
    \begin{subfigure}{0.48\columnwidth} 
        \includegraphics[width=\linewidth]{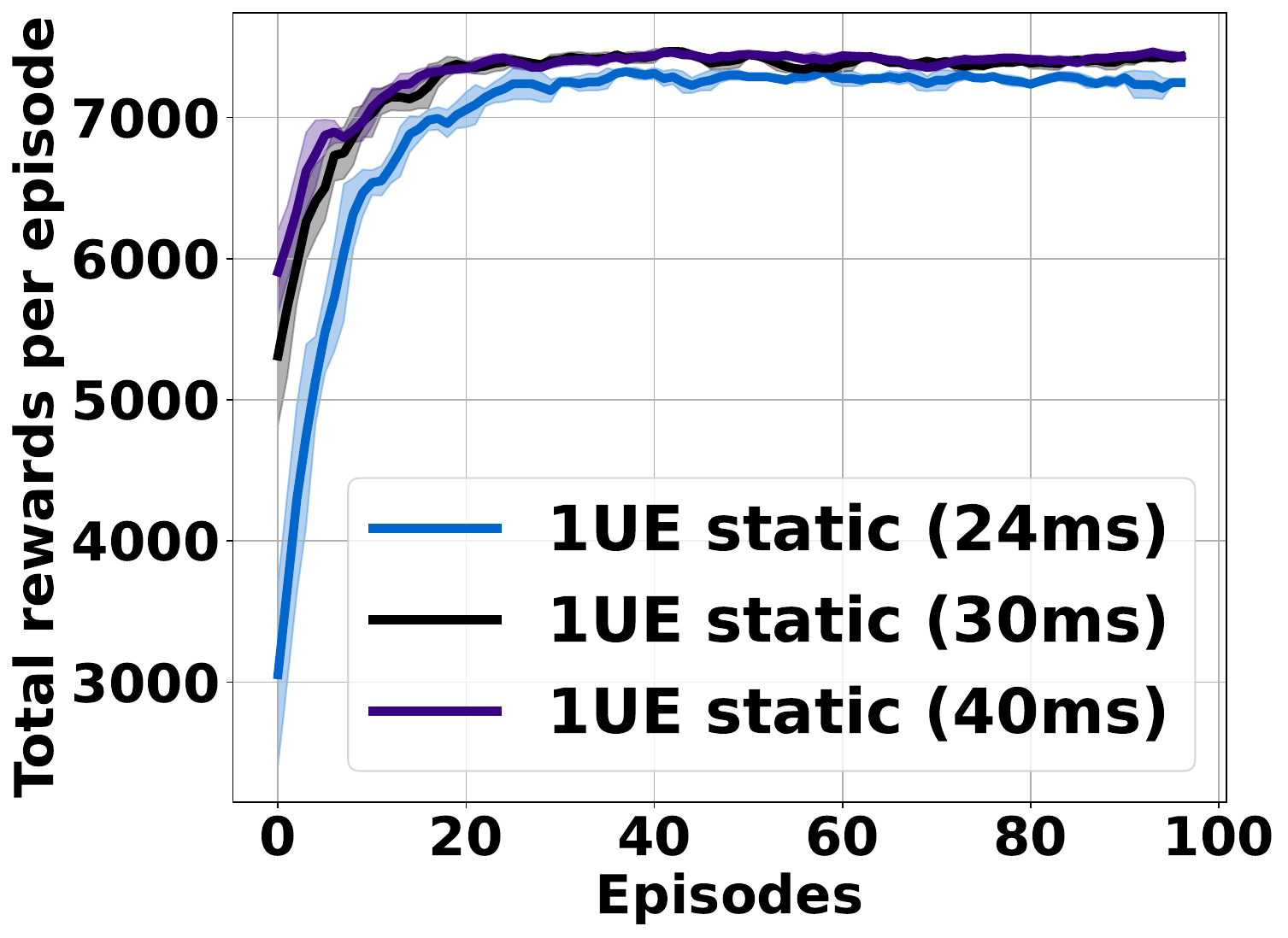}
        \caption{Single static UE.}
        \label{fig:static_ue_a}
    \end{subfigure}
    \hspace{1mm} 
    \begin{subfigure}{0.48\columnwidth} 
        \includegraphics[width=\linewidth]{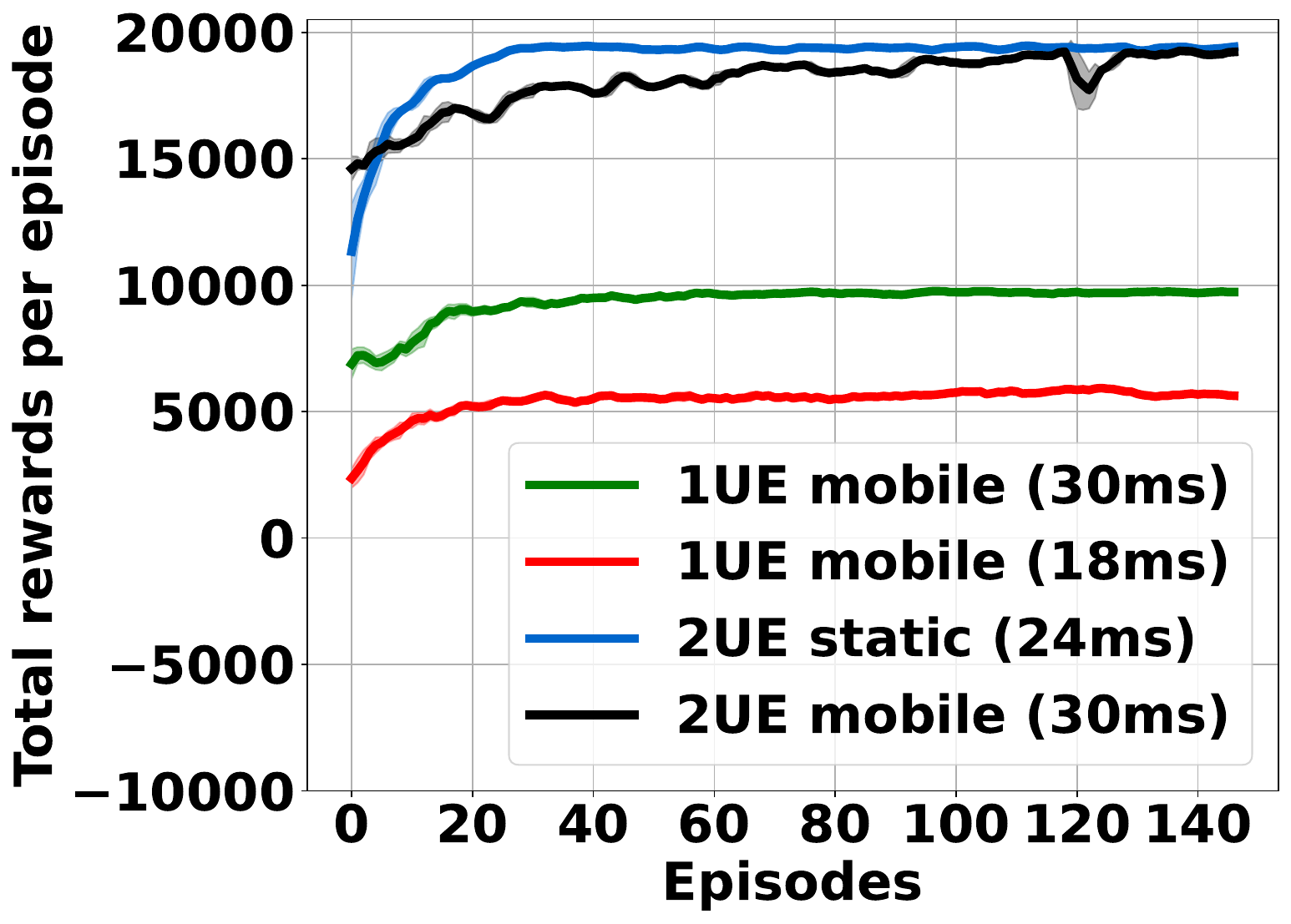}
        \caption{Multiple UEs, static\,\&\,mobile.}
        \label{fig:static_ue_b}
    \end{subfigure}
    \vspace{-5mm}
    \caption{Rewards achieved by the DRL agent for optimizing the network for different network setup scenarios considered.}
    \vspace{-3mm} 
    \label{fig:rewards}
\end{figure}

\begin{figure}[t!] 
    \centering
    \begin{subfigure}{0.48\columnwidth} 
        \includegraphics[width=\linewidth]{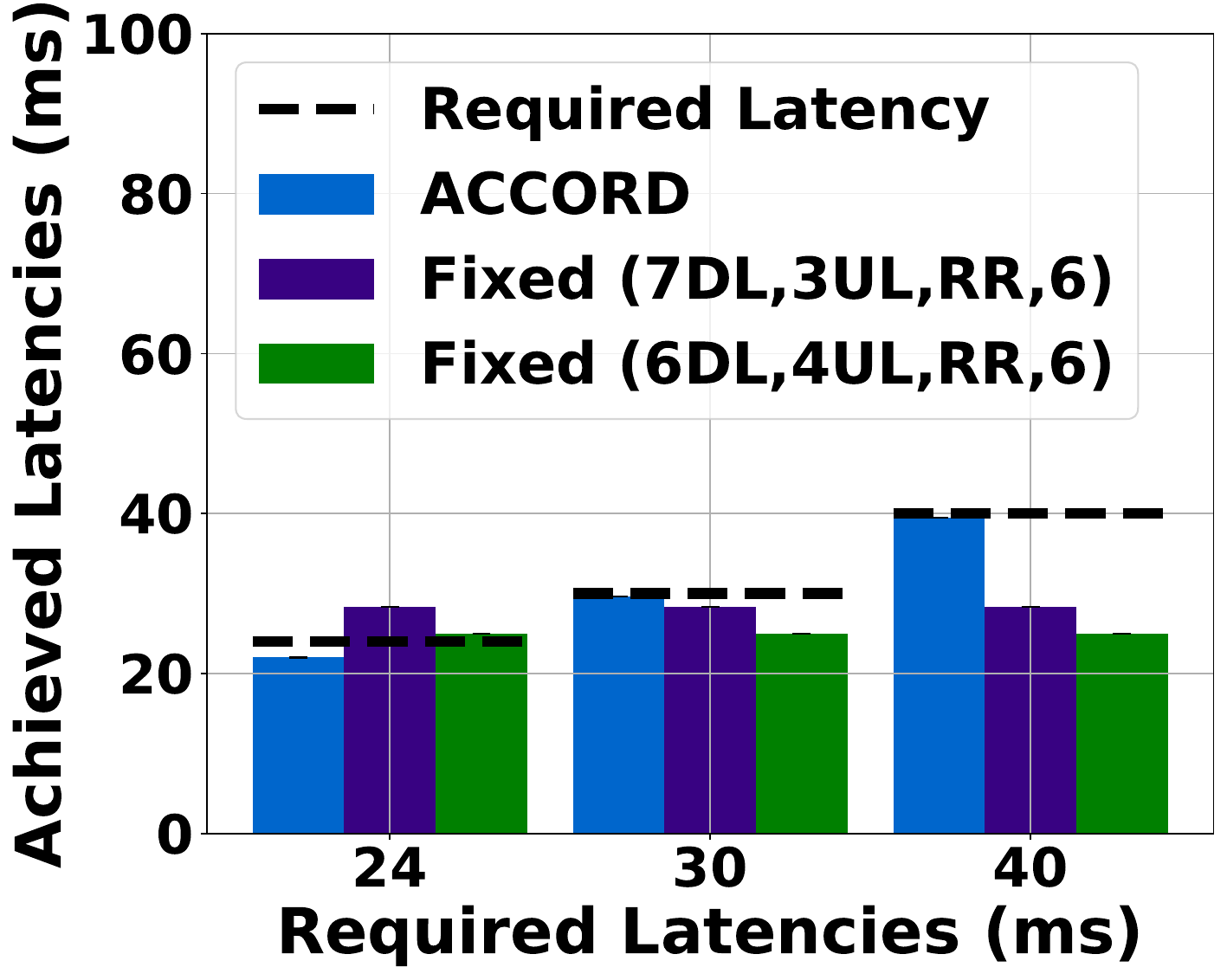}
        \caption{Single Static UE at a fixed location of 5000\,m from the gNB with a constant CQI of 10.}
        \label{fig:static_ues_a}
    \end{subfigure}
    \hspace{1mm} 
    \begin{subfigure}{0.48\columnwidth} 
        \includegraphics[width=\linewidth]{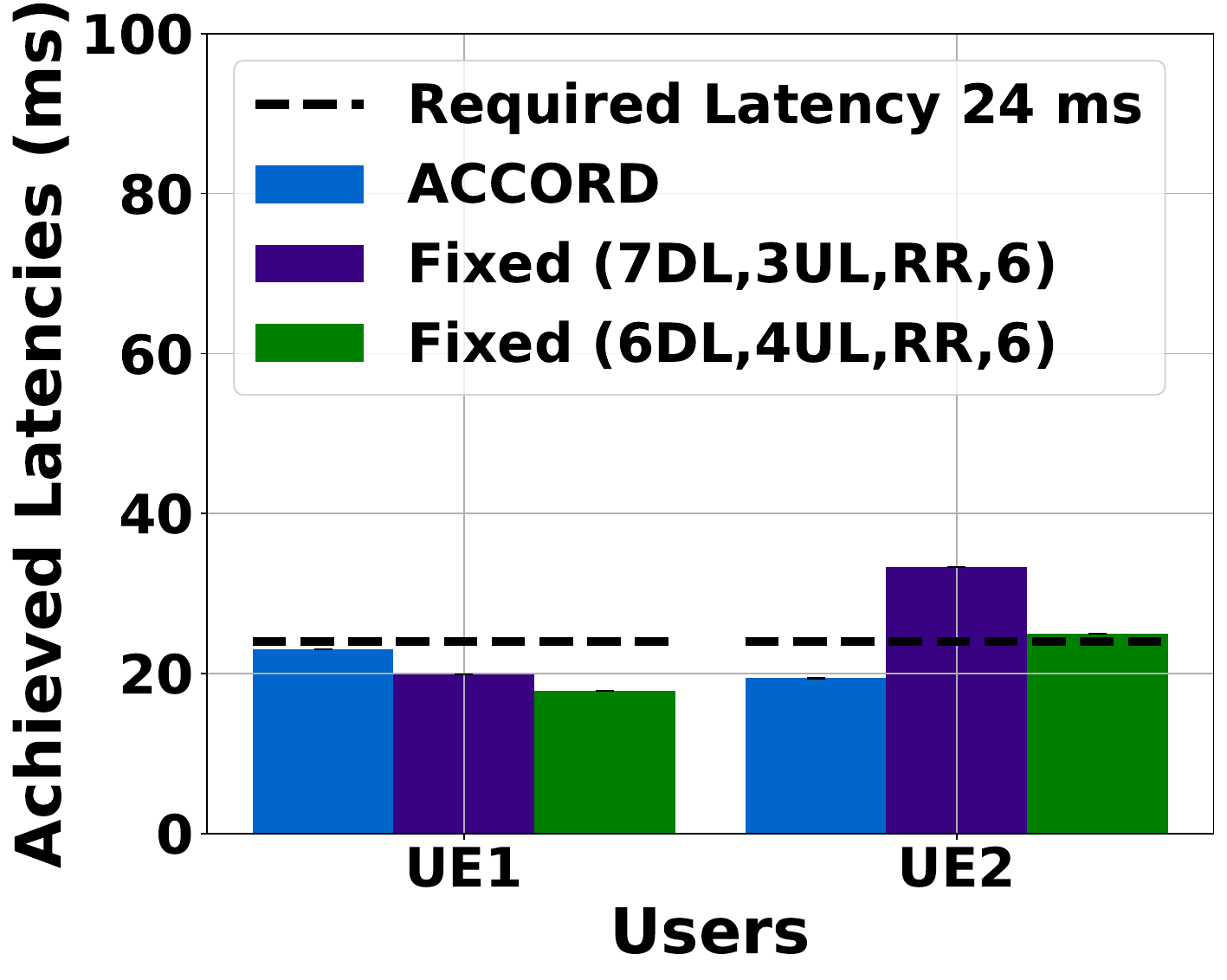}
        \caption{Two static UEs located 600m and 3600m from the gNB with CQI 15 and 8, respectively.}
        \label{fig:static_ues_b}
    \end{subfigure}
    \caption{Performance of various RAN configurations in meeting different latency requirements for static UEs.}
    \vspace{-3mm} 
    \label{fig:static_ues}
\end{figure}

\vspace{-0.2in}
\begin{figure*}[h!]
\centering
\begin{subfigure}{0.24\textwidth} 
    \includegraphics[width=\textwidth]{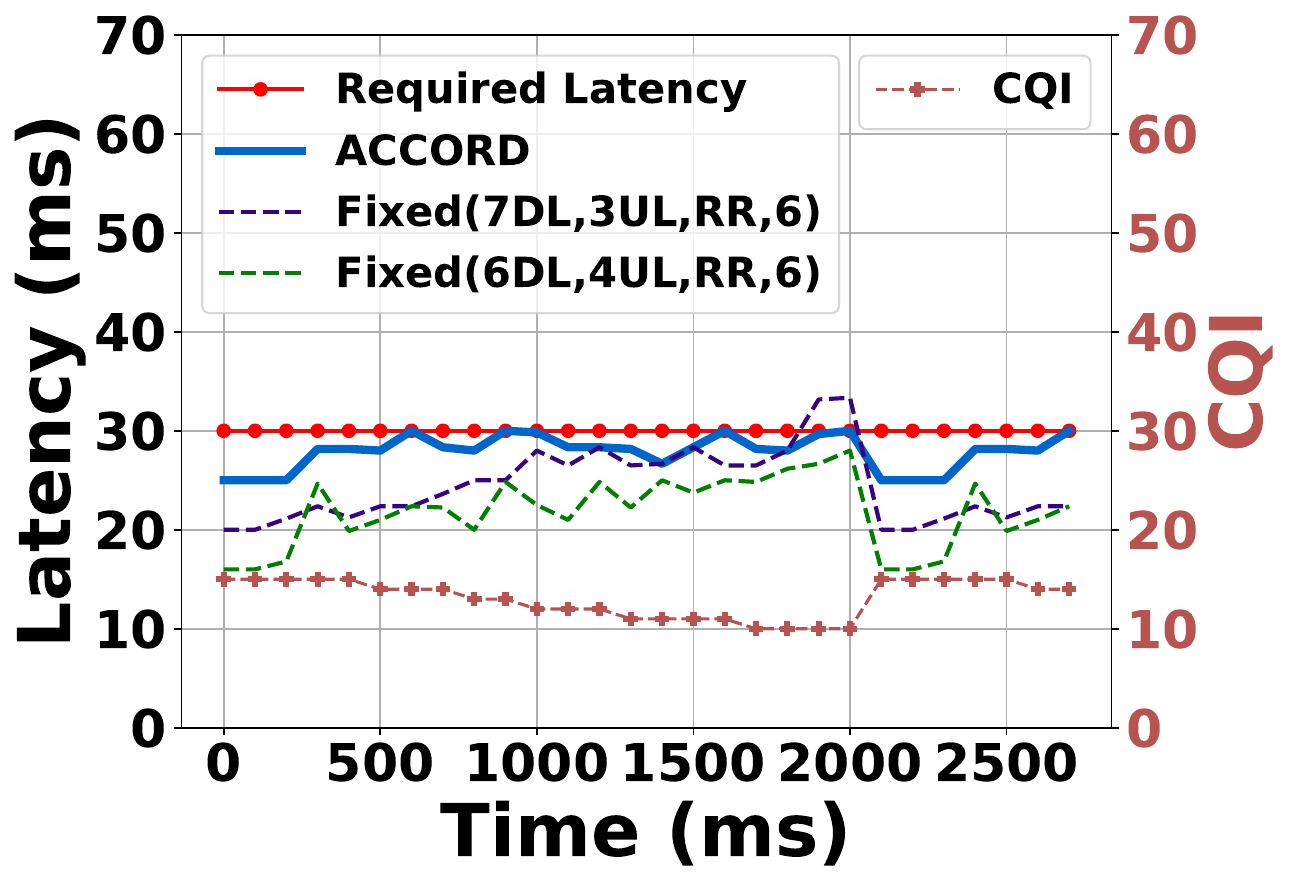}
    \caption{Latency \& CQI trends}
    \label{fig:1ue_time_cdf_plots_a}
\end{subfigure}\hfil 
\begin{subfigure}{0.22\textwidth} 
    \includegraphics[width=\textwidth]{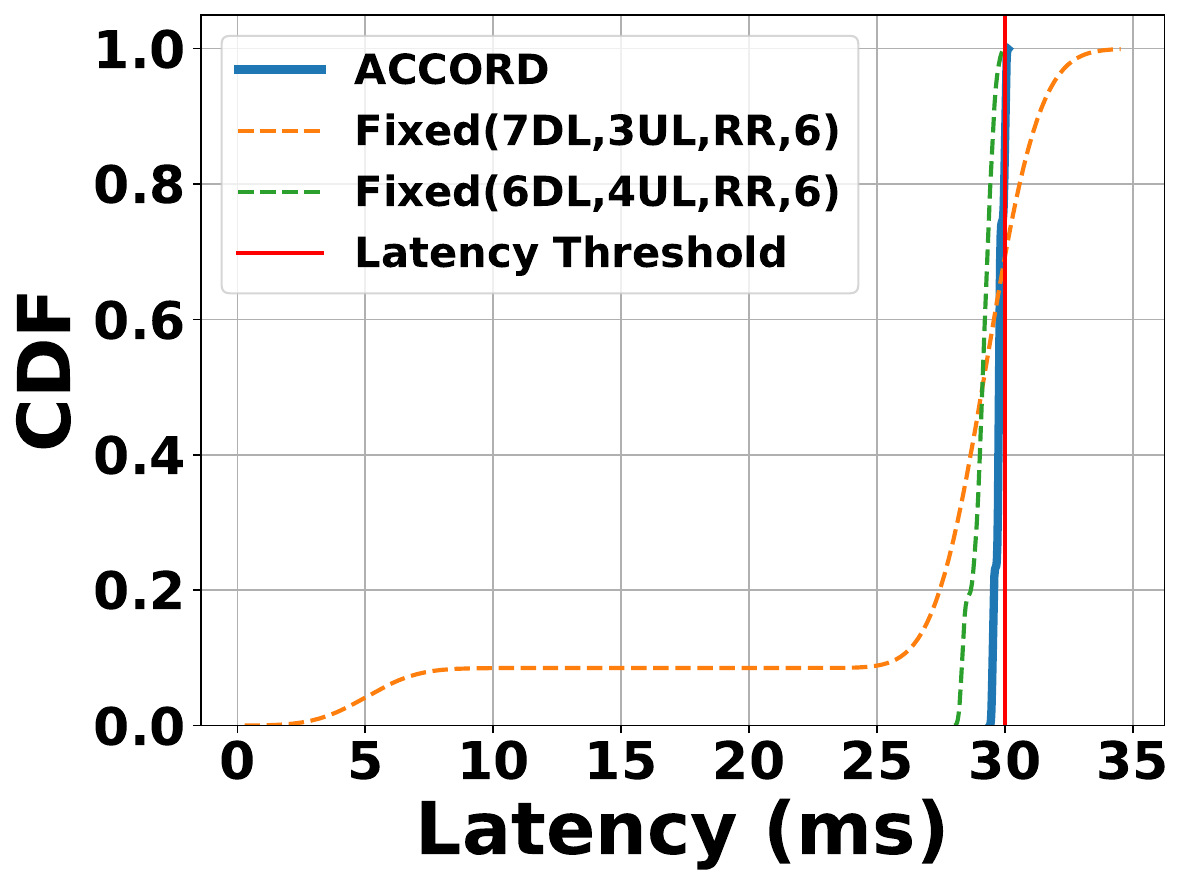}
    \caption{30\,ms latency}
    \label{fig:1ue_time_cdf_plots_b}
\end{subfigure}\hfil 
\begin{subfigure}{0.24\textwidth} 
    \includegraphics[width=\textwidth]{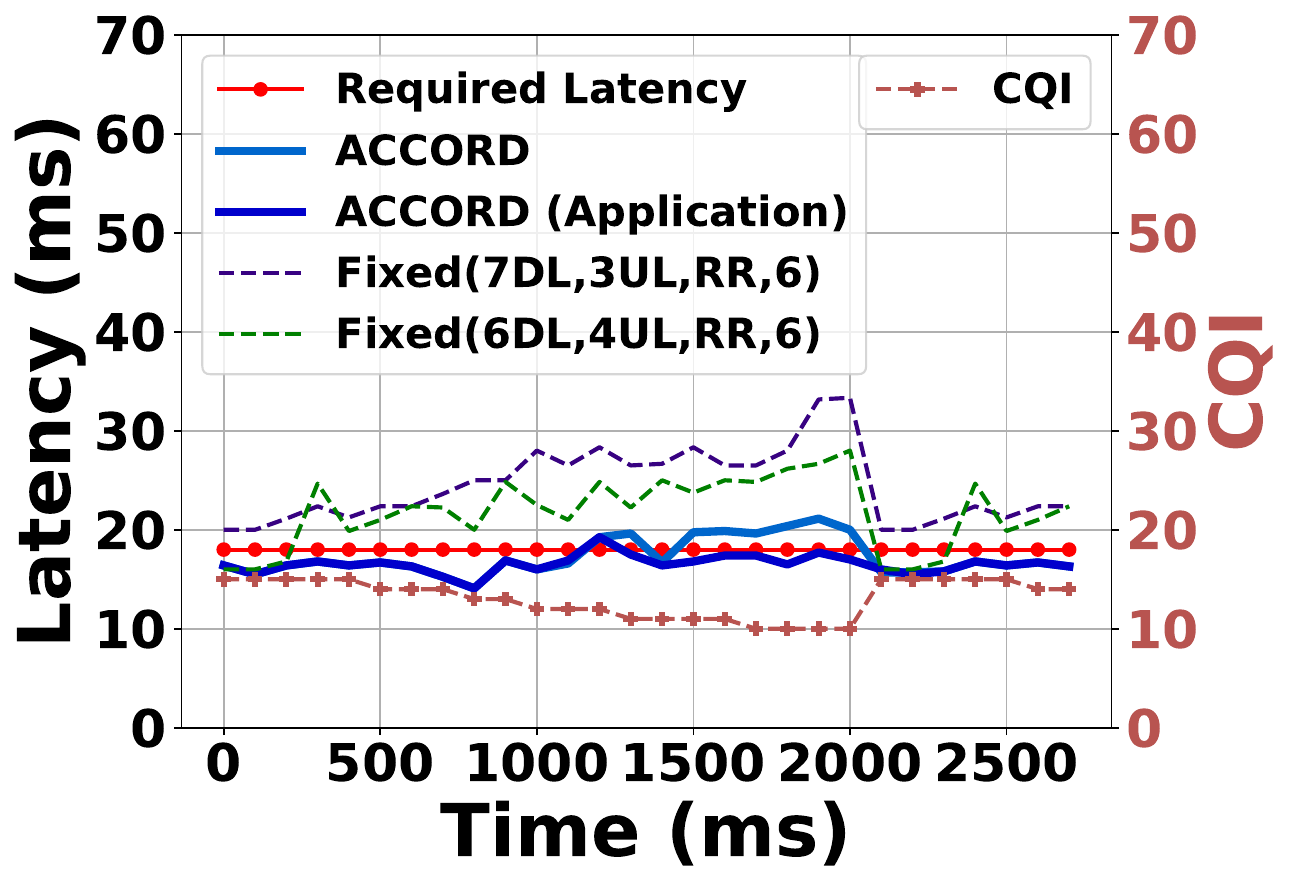}
    \caption{Latency \& CQI trends}
    \label{fig:1ue_time_cdf_plots_c}
\end{subfigure}\hfil 
\begin{subfigure}{0.22\textwidth} 
    \includegraphics[width=\textwidth]{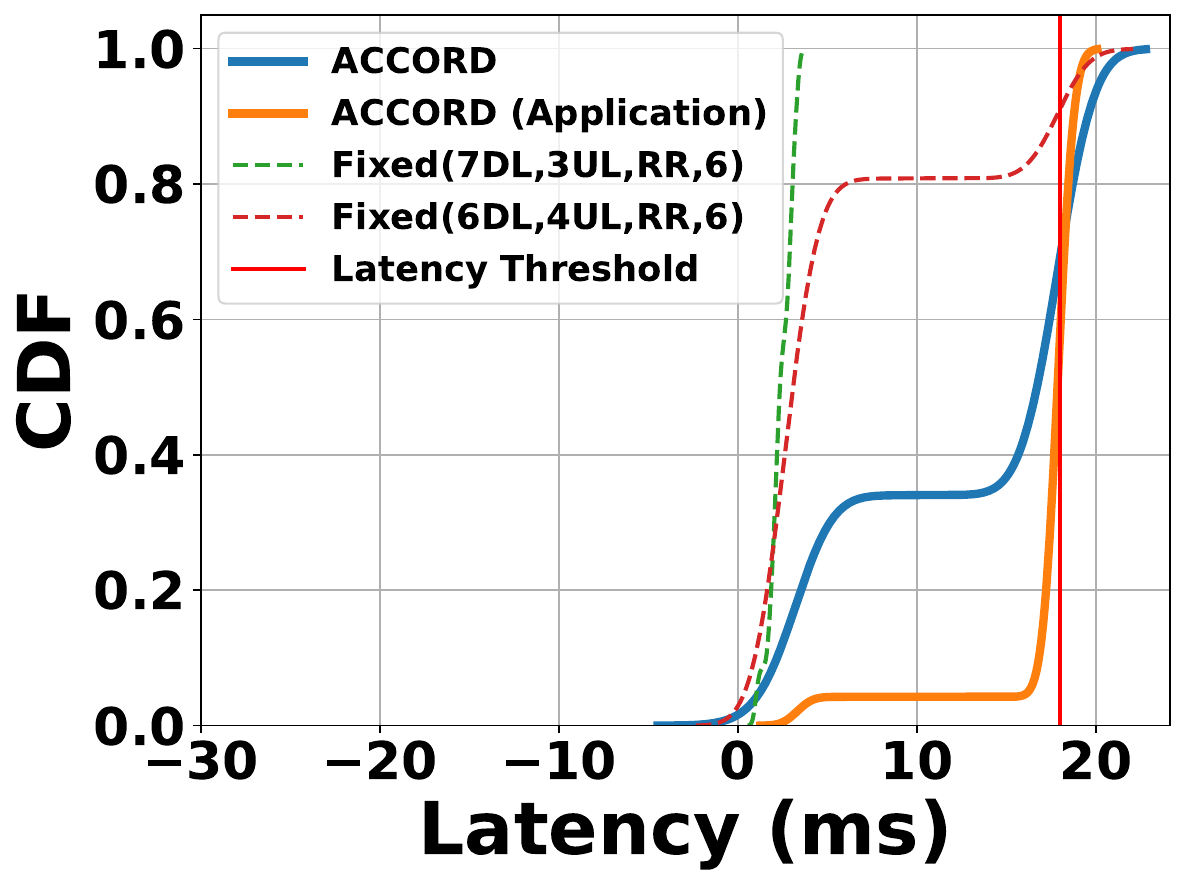}
    \caption{18\,ms latency}
    \label{fig:1ue_time_cdf_plots_d}
\end{subfigure}
\caption{Performance of various RAN configurations in meeting various latency requirements (30\,ms \& 18\,ms) for a mobile single UE with varying CQI over time (Emulating a UE moving away from gNB).}
\vspace{-2mm}
\label{fig:1ue_time_cdf_plots}
\end{figure*}
\begin{figure*}[h!]
\centering
\begin{subfigure}{0.24\textwidth} 
    \includegraphics[width=\textwidth]{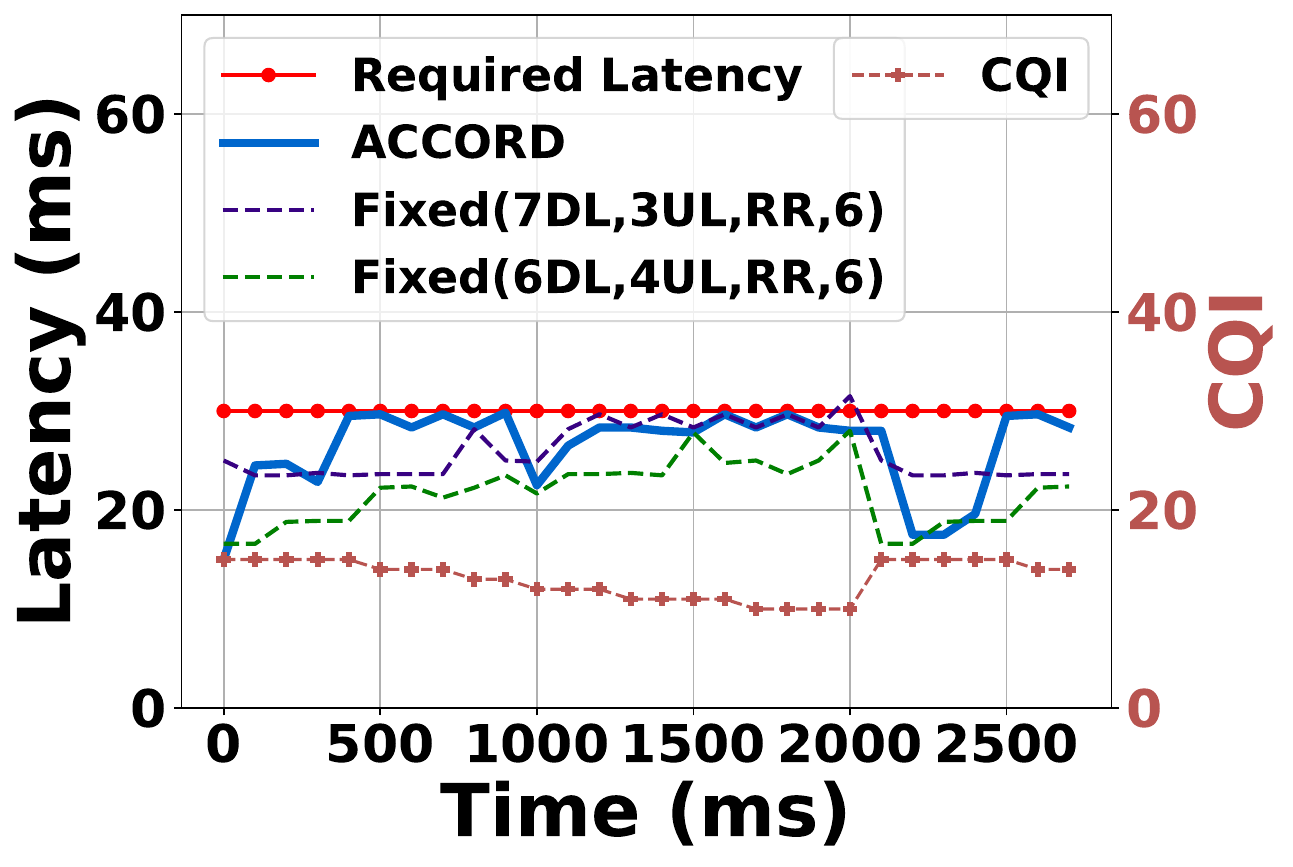}
    \caption{UE\,1 latency \& CQI trends.}
    \label{fig:2ue_time_cdf_plots_a}
\end{subfigure}\hfil
\begin{subfigure}{0.22\textwidth} 
    \includegraphics[width=\textwidth]{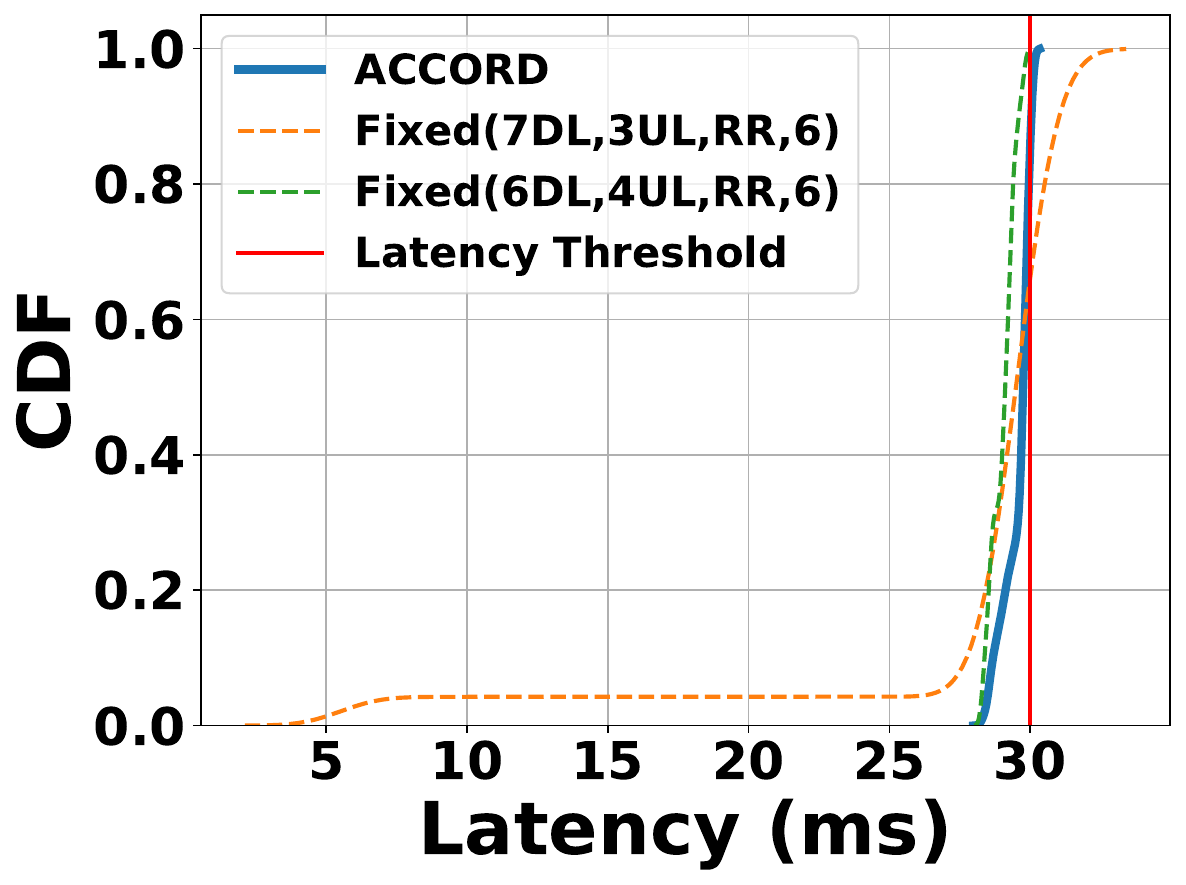}
    \caption{30\,ms latency (UE1).}
    \label{fig:2ue_time_cdf_plots_b}
\end{subfigure}\hfil
\begin{subfigure}{0.24\textwidth} 
    \includegraphics[width=\textwidth]{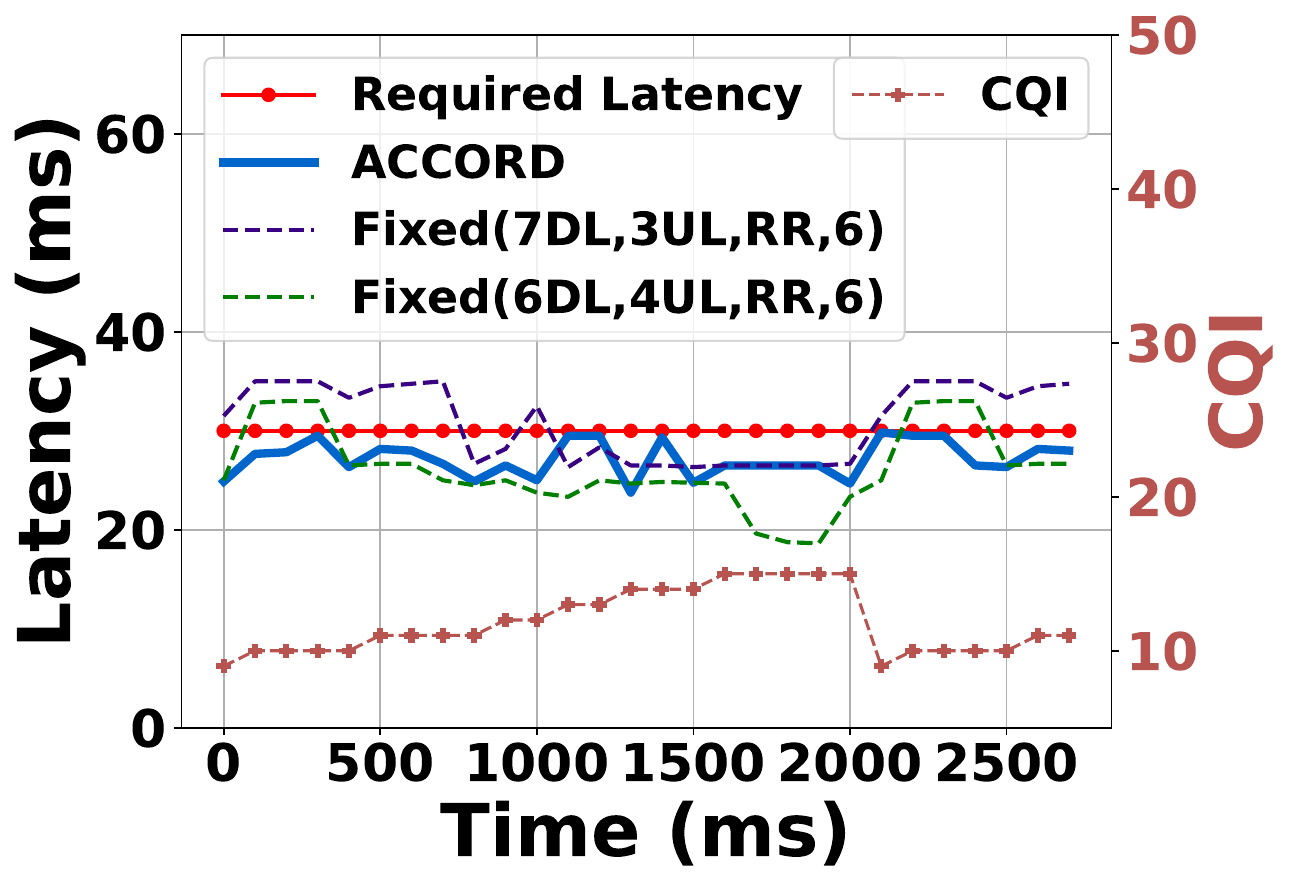}
    \caption{UE\,2 latency \& CQI trends.}
    \label{fig:2ue_time_cdf_plots_c}
\end{subfigure}\hfil
\begin{subfigure}{0.22\textwidth} 
    \includegraphics[width=\textwidth]{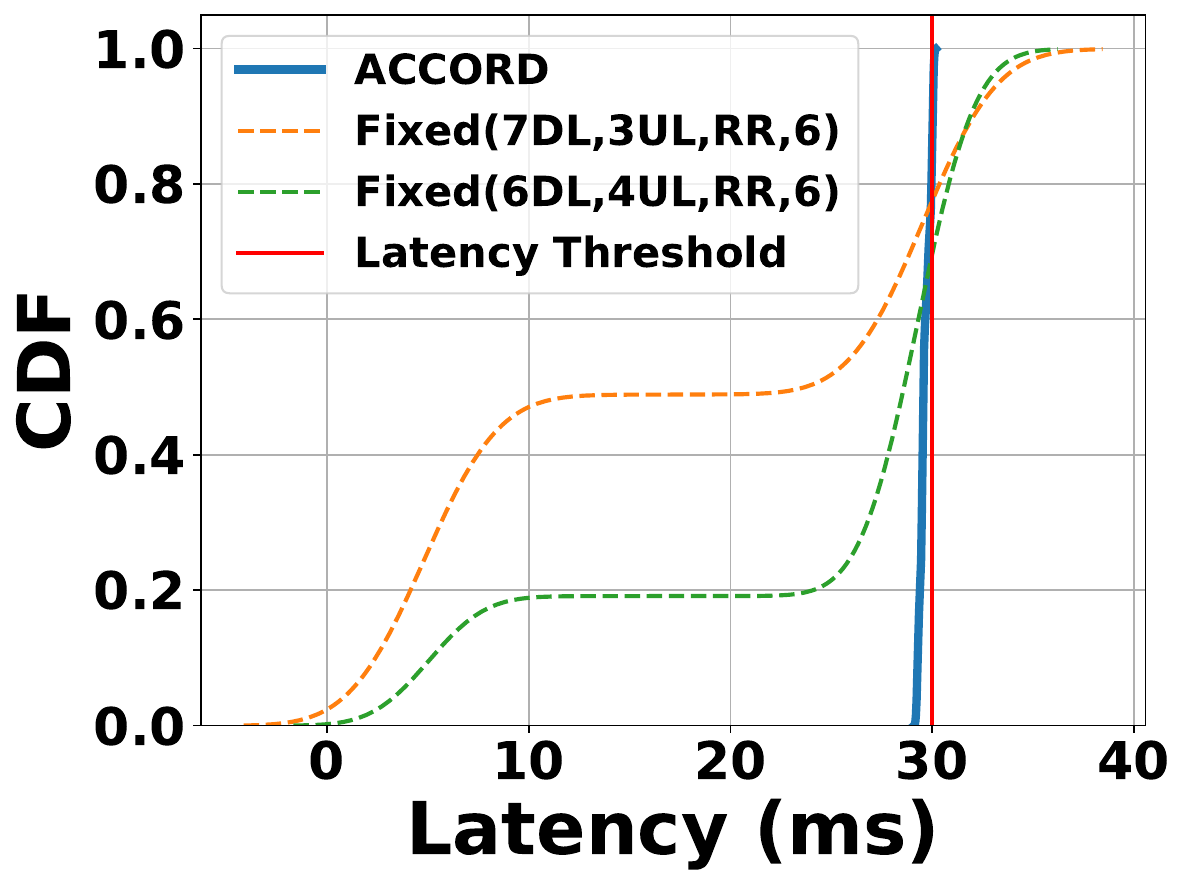}
    \caption{30\,ms latency (UE2).}
    \label{fig:2ue_time_cdf_plots_d}
\end{subfigure}
\caption{Performance of various RAN configurations for two mobile UEs in meeting a latency requirement of 30\,ms. Both UEs have varying CQI over time (UE\,1 emulates moving away from gNB while UE\,2 emulates moving towards gNB).}
\vspace{-4mm}
\label{fig:2ue_time_cdf_plots}
\end{figure*}
\vspace{0.2in}
\noindent $\bullet$ \textbf{Learning Performance: }Fig.\,\ref{fig:rewards} illustrates the learning progress of the DRL agent in each scenario, indicated by the total reward accumulated per episode during training. For the single stationary UE scenario (Fig.\,\ref{fig:static_ue_a}), the agent exhibits rapid convergence to the maximum reward, 
for all the considered uplink latency thresholds (24\,ms, 30\,ms, and 40\,ms). With 150 iterations per episode and a maximum reward of 50 per iteration, the achievable maximum reward per episode is 7500. Similarly, for the multiple UEs scenario (Fig.\,\ref{fig:static_ue_b}), the agent demonstrates faster convergence in the static case compared to the mobile case. This observation validates the effectiveness of the DRL agent in learning optimal policies for both single and multiple UEs.
To facilitate learning in the more challenging mobile scenario, we employed longer training episodes, allowing the agent more time to explore the state-action space and adapt to the dynamic channel conditions.\newline

{\em These results underscore the capability of the DRL agent to effectively learn and adapt to diverse network conditions, optimizing configuration parameters across multiple layers to meet the different latency requirements of MCAs across multiple UEs. The observed convergence behavior further suggests the robustness and stability of the DRL in ACCORD.}
\vspace{-0.01in}
\subsection{ACCORD Performance Results}
This section analyzes the performance of ACCORD in comparison to the fixed configuration approach of legacy 5G networks.\newline
\noindent $\bullet$ \textbf{Observation (Scenario A - Stationary UEs): }Fig.\,\ref{fig:static_ues_a} illustrates the achieved latency for a single static UE under varying latency requirements (24\,ms, 30\,ms, and 40\,ms). ACCORD consistently outperforms the fixed configurations of legacy 5G networks, demonstrating its ability to use information from the context to dynamically adjust network parameters and closely match the desired latency. This adaptability is crucial in resource-constrained environments, as the framework avoids over-provisioning resources and efficiently utilizes available bandwidth. For instance, when the required latency is 24\,ms, ACCORD converges to a configuration of 5\,DL-5\,UL slots, Round Robin scheduling, and RLC buffer size of 6\,KB, to meet the target latency. Conversely, for a 40\,ms latency requirement, the agent selects a less resource-intensive configuration of 6\,DL-4\,UL slots, Round Robin scheduling, and a buffer size of 2\,KB.  This adaptive behavior contrasts with the fixed configurations, which often overshoot or undershoot the latency target, leading to either over-allocation of network resources or performance degradation. Similar trends are observed in Fig.\,\ref{fig:static_ues_b} for the multiple static UEs scenario.  Despite the varying channel conditions and individual latency requirements of the two UEs, ACCORD successfully meets the target latency of 24\,ms for both, demonstrating its ability to handle diverse user demands and optimize network resources accordingly.\newline

\vspace{-0.1in}
\noindent $\bullet$ \textbf{Observation (Scenario B - Single Mobile UE): }Fig. \ref{fig:1ue_time_cdf_plots_a} depicts the performance of ACCORD and two fixed 5G configurations in meeting the latency requirement of a single mobile UE.
As the UE moves away from the gNB, its channel quality given by the CQI degrades.
While the fixed configurations can initially meet the latency requirement, their performance deteriorates as the CQI drops below 13. In contrast, ACCORD proactively adapts to the changing channel conditions and MCA context, dynamically adjusting network parameters to consistently maintain the desired latency. This adaptability is further evident in the statistical analysis presented in Fig.\,\ref{fig:1ue_time_cdf_plots_b}. The cumulative distribution function (CDF) of achieved latency demonstrates that ACCORD not only consistently meets the latency requirement but also operates close to the target without excessive over-provisioning of resources. For instance, ACCORD initially employs a configuration of 7\,DL-3\,UL slots, Proportional Fair scheduling, and a buffer size of 6\,KB at a CQI of 15. As the CQI degrades to 12 and the latency approaches the threshold, ACCORD incrementally increases the buffer size to 8\,KB and then 10\,KB.  With further CQI degradation, ACCORD switches to a new configuration with 6\,DL-4\,UL slots, Round Robin scheduling, and a buffer size of 2\,KB. This dynamic adaptation highlights ACCORD's ability to effectively utilize network resources and respond to changing channel conditions. Fig.\,\ref{fig:1ue_time_cdf_plots_c} examines a more stringent latency requirement, where the fixed configurations struggle to meet the target as the UE moves away from the gNB.  This scenario underscores the limitations of fixed configurations in dynamic environments, particularly with limited bandwidth (5\,MHz). To address this, the agent optimizes APP layer parameters, enabling it to consistently meet the latency requirement even with significant CQI degradation.

\noindent $\bullet$ \textbf{Observation (Scenario B - Multiple Mobile UEs):} Similar to the single mobile UE case, ACCORD outperforms the fixed 5G configurations in the multiple mobile UE scenario, effectively handling the complexities introduced by varying channel conditions and individual UE mobility patterns.\newline
Fig.\,\ref{fig:2ue_time_cdf_plots} illustrates the performance analysis for a scenario with two mobile UEs experiencing dynamic channel conditions due to their movement relative to the gNB. ACCORD successfully meets the latency requirements of both UEs by selecting unique configurations at the RLC layer while maintaining common configurations across the PHY and MAC layers.  This differentiation stems from the varying CQI levels experienced by each UE, with one moving away from the gNB (degrading CQI) and the other moving towards it (improving CQI).  Applying a uniform configuration across both UEs would fail to address their individual needs and maintain consistent latency performance.  ACCORD intelligently adjusts the RLC layer buffer size to compensate for the varying channel conditions, ensuring that both UEs meet their respective latency requirements despite the dynamic environment.  This result underscores the importance of learning and responding to individual UE channel conditions in a multi-user scenario.\newline
These results collectively demonstrate the efficacy of ACCORD in optimizing 5G network performance for mobile applications. The ability to dynamically adapt to changing channel conditions, efficiently utilize network resources and optimize application layer parameters positions it as a powerful solution for ensuring QoS with efficient spectrum utilization in dynamic and demanding 5G environments.

{\em Overall, ACCORD's ability to differentiate configurations at the RLC layer while maintaining common parameters at the PHY and MAC layers highlights its nuanced and efficient resource allocation capacity, enabling the framework to effectively manage diverse user demands and optimize network performance even in complex and dynamic environments.}

\section{Conclusion}
This research demonstrates the effectiveness of ACCORD, a context-aware DRL framework for optimizing 5G resource allocation in machine-centric applications, compared to legacy 5G solutions. ACCORD uses application requirements supplemented by contextual information as input to a DQN agent to meet event-driven latency requirements. This agent learns to dynamically adjust network parameters across the PHY, MAC, RLC, and application layers, adjusting to the needs of individual devices and channel conditions. The framework's ability to increase spectrum efficiency and deliver consistent latency performance was validated by evaluations conducted in various scenarios. For future work, ACCORD will be generalized for heterogeneous applications, real-world 5G deployments, and optimization across extra network layers.
\bibliographystyle{IEEEtran}

\bibliography{ref}

\end{document}